\documentstyle[twocolumn,aps,epsf]{revtex}
\begin{document}
\wideabs{

\title{Self-trapping transition for nonlinear impurities embedded in a 
       Cayley tree}

\author{Bikash C. Gupta and Sang Bub Lee}
\address{Department of Physics, Kyungpook National University,
Taegu 702--701, Korea}

\maketitle

\begin{abstract}
The self-trapping transition due to a single and a dimer nonlinear
impurity embedded in a Cayley tree is studied. 
In particular, the effect of a perfectly nonlinear Cayley tree is considered. 
A sharp self-trapping transition is observed in each case. 
It is also observed that the transition is much sharper compared 
to the case of one-dimensional lattices. 
For each system, the critical values of $\chi$ for the self-trapping 
transitions are found to obey a power-law behavior as a function of 
the connectivity $K$ of the Cayley tree.
\end{abstract}
}

\section{Introduction}
The interaction of an electron or an exciton with the lattice vibrations
is of fundamental importance in understanding the electric 
properties of solids.
For example, the transport of quasi-particles such as electrons 
or excitons in solids is highly influenced by the electron-phonon interactions. 
The consequences have been investigated using different 
methods\cite{1}
Recently, these systems have been studied based on the rigorous 
analytical treatments and numerical solutions of simple models 
such as nonlinear Schr\"odinger equations\cite{2,3}. 
One of them with varieties of applications in different areas of science 
is the one-dimensional discrete nonlinear Schr\"odinger equation,
given as\cite{4,5,6,7,8,9,10,11,12,13,14,15,16,17,18}
\begin{equation}
i\frac{dC_n}{dt}=V(C_{n+1} + C_{n-1}) + (\epsilon_n - \chi_n |C_n|^2) C_n ,
\end{equation}  
where $\epsilon_n$ is the static site energy at site $n$ and $\chi_{n}$ is 
the nonlinearity parameter associated with the $n$th grid point. 
Since $\sum_{n} \mid C_n \mid^2$ is set to be unity by choosing 
appropriate initial conditions, 
$\mid C_n \mid^2$ can be considered as a probability of finding a particle 
at the $n$th grid point. 
One way to derive this set of equations is to couple in the 
adiabatic approximation (in which the lattice oscillations are much faster
than the exciton motion) the vibration of masses at lattice points 
of a lattice of $N$ sites to the motion of a quasi-particle in
the same lattice. 
The motion of a quasi-particle is described, however, in the frame work of a 
tight binding Hamiltonian (TBH).
In other physical context,  the set of equations are often called the 
discrete self-trapping equations (DST). 

The analytical solutions of Eq.~(1) are, in general, not known.
However, for nonlinear quantum dimers which are two-site systems with
the nonlinearity either on both the site-energies or in one of them can
be solved analytically for any arbitrary initial condition. 
>From the analytical solutions, a self-tapping transition is found in this
model\cite{6,7,8,9,10}. 
The self-trapping transition for the symmetric dimer is found 
at $\chi/V$=4\cite{6}. 
The trapping of hydrogen ions surrounding oxygen atoms in metal hydrides 
and the energy transport from the absorption center to the reaction center 
in photosynthetic unit have been modeled by the effective quantum nonlinear 
dimer\cite{6,7,8,9,10,12,13,17}. 
The nonlinear dimer analysis has also been applied to several 
experimental situations like neutron scattering of hydrogen atoms 
trapped at the impurity sites in metals\cite{7}, 
fluorescence depolarization\cite{9}, muon spin relaxation\cite{10}, 
nonlinear optical response of superlattices\cite{19}, etc.
The self-trapping transition also occurs in the extended nonlinear
systems\cite{5,17}. 
An interesting experimental example in this context is the observation 
that trapped hydrogen atoms in metals like Nb move among the sites 
in the neighborhood of impurity atoms such as oxygen\cite{20}.
All these studies have been performed for finite number of nonlinear sites 
by assuming that the quasi-particle is localized within the nonlinear sites.  

Later, the effect of nonlinear sites embedded in a host
lattice on the dynamics of quasi-particles has been studied
because of its importance in real systems. 
Dunlap {\em et al.}\cite{21} studied the self-trapping transition 
due to a single nonlinear impurity embedded in an infinite one-, 
two- and three-dimensional host lattices. 
Self-trapping transitions were found at $\chi/V$=3.2, 6.72 and 
9.24 for one-, two- and three-dimensional simple host lattices respectively. 
Furthermore, the effect of the presence of a nonlinear cluster on
the self-trapping transition has also been considered in one-dimensional 
host lattices\cite{datt}. 
The study has also been extended to the case where the inertial effect 
of the lattice oscillators has been taken into account and rich 
trapping-detrapping transitions depending on the masses of oscillators 
have been observed\cite{22}. 
All these studies have been made in one-dimensional host lattices. 
However, one needs to know whether or not the self-trapping transition occurs 
in hosts of different geometrical structure. 
It would also be interesting to note the differences compared to the
results for one-dimensional systems.

The Cayley tree is one possible example of host lattices with 
different geometrical structure. 
An important variable characterizing the geometry of the Cayley tree is 
the connectivity $K$, which is the number one smaller than the 
coordination number, i.e., $K=Z-1$, $Z$ being the coordination number. 
The Cayley tree reduces to a one-dimensional chain if $K=1$.
The structure of the Cayley tree will be described in the next section. 

In this work, we study the self-trapping of an exciton in the Cayley tree 
with a single as well as a dimer impurity embedded in it. 
We also consider the fully nonlinear Cayley tree to observe 
the self-trapping effect.  

\section{Formalism}
The structure of a Cayley tree with connectivity $K=2$ is shown in Fig.~1.
For $K=1$, the system reduces to a one-dimensional chain. 
In case of a single nonlinear impurity embedded in the Cayley tree, 
the neighboring sites surrounding the impurity site are symmetric. 
We thus consider the symmetric shells around the impurity site. 
The impurity site is designated as the zeroth site, 
and the $n$th shell is denoted by $n$, 
where $n$ takes the values of 1, 2, 3, $\cdots$, 
as the sites are away from the zeroth site (see Fig.~1). 

\vspace{1.0cm} 
\begin{figure}
  \begin{center}
  \leavevmode
  \epsfxsize=7.0cm
  \centerline{\epsfbox{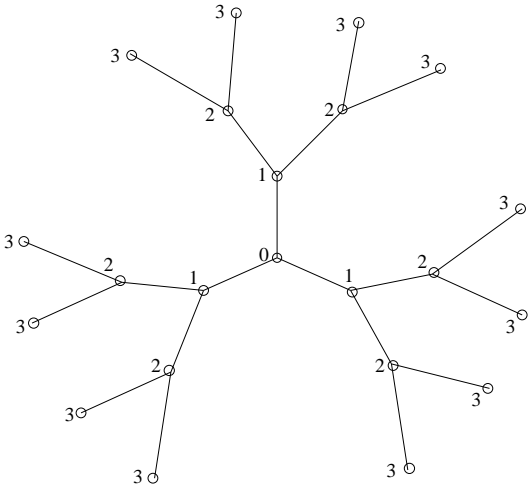}}
  \vspace{0.5cm} 
  \caption{The Cayley tree of the connectivity $K=2$.
           The impurity is embedded at the site marked by 0.
           The sites in the first shell are numbered as 1,
           the sites in the second shell as 2, and so on.}
  \label{fig1}
  \end{center}
\end{figure}

Thus, the $n$th shell contains $ZK^{n-1}$ sites. 
All sites in the lattice have three nearest neighbors. 
While all the nearest-neighbor sites of the zeroth site fall in the first shell,
two of the nearest-neighbor sites of any site in the $n$th shell 
fall in the $(n+1)$th shell and the rest one falls in the $(n-1)$th shell.
We further notice that all sites in the same shell have an equal probability
amplitude.
Therefore, under the tight-binding formalism, the time evolution of a 
particle (initially placed at the impurity site) on the Cayley tree 
may be governed by the following equations:
\begin{eqnarray}
i\frac{dC_0}{dt}&=&Z C_1 - \chi |C_0|^2 C_0 \nonumber \\
i\frac{dC_n}{dt}&=&K C_{n+1} + C_{n-1},~~~n \ge 1
\end{eqnarray}
where $C_0$ is the probability amplitude at the zeroth site and $C_n$ for 
$n \ge 1$ represents the probability amplitude at any site in the $n$th shell. 
The $\chi$ represents the nonlinear strength at the
zeroth site of the Cayley tree. 
Without loss of generality, we take the nearest-neighbor hopping element 
to be unity. 
The normalization condition for the site amplitudes is given by
\begin{equation}
|C_o|^2 + \sum_{n=1}^\infty Z K^{n-1} |C_n|^2 = 1.
\end{equation}
Therefore, to observe the self-trapping transition due to a single
nonlinear impurity, Eq.~(1) should be solved.

For the perfectly nonlinear Cayley tree, 
the time evolution for the site amplitudes on the Cayley tree of a particle 
(initially placed at the zeroth site) may be governed by 
\begin{eqnarray}
i\frac{dC_0}{dt} &=& Z C_1 - \chi |C_0|^2 C_0 \nonumber \\
i\frac{dC_n}{dt} &=& K C_{n+1} + C_{n-1} - \chi |C_n|^2 C_n,~~~n \ge 1
\end{eqnarray}
because the system may be treated in a similar way as it is treated for
the single impurity case.  

\vspace{1.0cm}
\begin{figure}
  \begin{center}
  \leavevmode
  \epsfxsize=7.0cm
  \centerline{\epsfbox{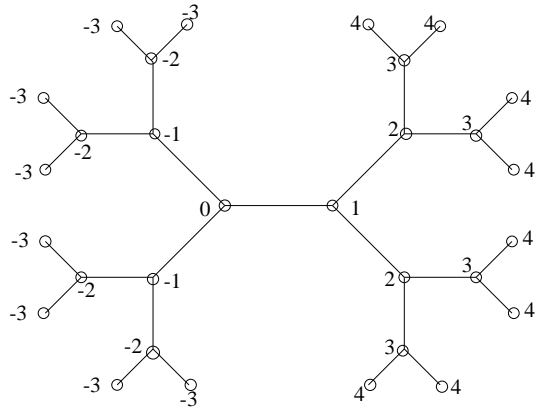}}
  \vspace{0.5cm}
  \caption{The Cayley tree with the connectivity $K=2$.
           The dimeric impurity is embedded at the sites marked by 0 and 1.}
  \label{fig2}
  \end{center}
\end{figure}

If there is a dimeric impurity in the system, the symmetry about 
one of the impurity sites does not hold anymore. 
The system, however, remains symmetric about the bond connecting the
two impurity sites. 
In this case, the Cayley tree with dimeric impurity may be transformed 
to a one-dimensional system which might be studied more conveniently.
The transformation has been reported in the earlier work\cite{Kundu+Gupta:1998};
however, for completeness, we briefly describe it in what follows. 

We pick a bond and assign the numbers 0 and 1 on its two ends.
The neighboring sites of the site 1 are numbered with an increasing order 
and those of the site 0 are numbered with a decreasing order, 
as shown in Fig.~2.
Thus, all points with the same number are assumed to be in the same generation
and, accordingly, the number of points in the $n$th generation is 
$K^{n-1}$ if $n \ge 1$ and $K^{|n|}$ if $n \le 0$. 
We note that all sites in a given generation have the same 
probability amplitude.

We now consider the motion of a particle on a Cayley tree of the
connectivity $K$ with a dimeric impurity embedded at sites 0 and 1. 
In the tight-binding formalism with nearest-neighbor hopping only,
equations governing the motion of a particle are 
\begin{eqnarray}
i \frac{d{C_n}}{d{t}}&=&K  C_{n+1} + 
C_{n-1}, ~~~~~~~~~~~n > 1 
\nonumber \\
i \frac{d{C_n}}{d{t}}&=&K C_{-|n|-1} + 
C_{-|n|+1},~~~~~n < 0 
\nonumber \\
i \frac{d{C_1}}{d{t}}&=&K C_{2} + 
C_{0} + \epsilon_1 C_1 \nonumber \\ 
i \frac{d{C_0}}{d{t}}&=&K C_{-1} + 
C_{1} + \epsilon_0 C_0 ,
\end{eqnarray}
where $C_n$ denotes the probability amplitude at any point
in the $n$th generation. 
We assume that the matrix element of the nearest-neighbor hopping is unity,
and that all points in a given generation arising due to a 
specific organization have the same site energy.
The normalization condition for the site amplitudes gives
\begin{equation}
\sum_{-\infty}^0 K^{|n|} |C_n|^2 + \frac{1}{K} \sum_{n=1}^\infty
K^n |C_n|^2 = 1.
\end{equation}

We now make the following transformations:
(i) $\tau = \sqrt K t$, 
(ii) $\epsilon_0 = \widetilde \epsilon_0 / \sqrt K$ and $\epsilon_1 
= \widetilde \epsilon_1 / \sqrt K$, 
(iii) $C_n = K^{-(n-1)/2} \widetilde C_n$ for $n \ge 1$ and 
$C_{-|n|} = K^{-|n|/2} \widetilde C_n$ for $n \le 0$.
Substituting these in Eq.~(5) we obtain 
\begin{eqnarray}
i \frac{d \widetilde C_n}{d\tau}&=&\widetilde C_{n+1} + \widetilde C_{n-1} 
, ~~{\rm for}~~ n >1 
~~{\rm and}~~ n < 0. \nonumber \\
i \frac{d\widetilde C_1}{d\tau}&=&\widetilde C_{2} + \frac{1}{\sqrt K}
\widetilde C_{0} + \widetilde \epsilon_1 \widetilde C_1, 
\nonumber \\ 
i \frac{d\widetilde C_0}{d\tau}&=&\widetilde C_{-1} + \frac{1}{\sqrt K}
\widetilde C_{1} + \widetilde \epsilon_0 \widetilde C_0. 
\end{eqnarray}
>From Eq.~(7), the normalization condition reduces to
$\sum_{-\infty}^{\infty} |\widetilde C_n|^2$ = 1. 
Therefore, the motion of a particle on a Cayley tree is mapped to 
that on a one-dimensional chain. 
However, in this chain, the nearest-neighbor hopping matrix element 
between the zeroth and first site is reduced from unity to $1/\sqrt K$. 
It can be shown that the Green's function $G_{0,0}(E)$ calculated from Eq.~(7)
would yield $\widetilde G_{0,0} (E = \widetilde E\sqrt K)$ 
for a Cayley tree of the connectivity $K$.
Here, in the dimeric impurity problem, the impurities are defined
as $\widetilde \epsilon_0 = \widetilde \chi |C_0|^2$ and $\widetilde 
\epsilon_1 = \widetilde \chi |C_1|^2$ with $\widetilde \chi=
\frac{\chi}{\sqrt K}$. 

Eqs.~(2), (4) and (7) can not be solved analytically and, therefore,
the numerical method, namely, the fourth order Runge Kutta method is employed. 
Since there is a conserved quantity in each case, 
the normalization condition is checked at every step of 
our numerical calculation. 
The time interval $\delta t =0.001$ is used during the calculation.
There are two ways to observe the self-trapping transition;
One way is to look at the behavior of $|C_n|^2$ as a function of $t$ 
for various values of the nonlinear strength, and another is to look at the 
time averaged probability of the particle at site $n$ as a function of 
the nonlinear strength.
The time averaged probability of the exciton at site $n$ is defined as
\begin{equation}
<P_n> = \lim_{T \rightarrow \infty} \frac{1}{T} \int_0^T |C_n|^2 dt.
\end{equation}
Therefore, we will look at the quantity $|C_{n}|^2$ or $< P_{n} >$ or
both of them for various situations to examine the occurrence of the 
self-trapping transition.

\section{results and discussions}
First of all, we discuss the results for the Cayley tree with a 
single nonlinear impurity. 
The initial condition is set at the impurity site (zeroth site in Fig.~1). 

\vspace{0.7cm}
\begin{figure}
  \begin{center}
  \leavevmode
  \epsfxsize=7.5cm
  \centerline{\epsfbox{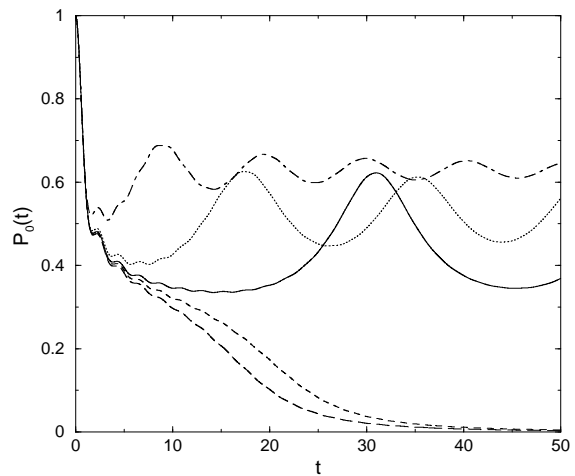}}
  \vspace{0.1cm}
  \caption{The probability of the particle to be at the impurity site
           ($P_{0} (t)$) as function of time for various values of nonlinear
           parameter is shown. The long dashed curve, dashed curve,
           solid curve, dotted curve and the dotted-dashed curve correspond,
	   respectively, to $\chi$ = 4.75, 4.76, 4.77, 4.8 and 5.0.}
  \label{fig3}
  \end{center}
\end{figure}

The probability of finding a particle at the impurity site 
(initially populated site) is obtained by solving Eq.~(2), 
and the results are plotted in Fig.~3 as a function of time. 
Different curves correspond to the different values of the nonlinear
parameter $\chi$. 
The long dashed, dashed, solid, dotted, and dotted-dashed curves
correspond to the nonlinear strength of $\chi$ = 4.75, 4.76, 4.77, 4.8 and 5.0,
respectively. 
The connectivity for the Cayley tree considered here is $K=2$. 

It is observed from Fig.~3 that, for $\chi$ = 4.75 and 4.76 
(the long-dashed and dashed curves), the probability of finding the particle 
at the impurity site decreases rapidly and then approaches to zero 
as time increases. 
This implies that the particle goes away from the impurity site, i.e., 
the particle becomes fully delocalized. 
However, the situation is drastically different for $\chi$ = 4.77 (solid curve).
The probability of finding the particle at the initially populated site 
decreases down to about 0.35, then increases up to 0.61, and oscillates
afterward between 0.35 to 0.61. 
Thus, the average probability of finding the particle at the initially 
populated site is approximately 0.48. 
For higher value of $\chi$, the probability of finding the particle 
at the initially populated site increases as is obvious from the dotted 
and dotted-dashed curves in Fig.~3. 
Therefore, we observe that there is a distinct critical value of 
$\chi$ near (or below) 4.77, below which the particle escapes from the 
initially populated site and becomes fully delocalized, while above which
the particle is most likely trapped at the initially populated site. 

\vspace{0.7cm}
  \begin{figure}
  \begin{center}
  \leavevmode
  \epsfxsize=7.5cm
  \centerline{\epsfbox{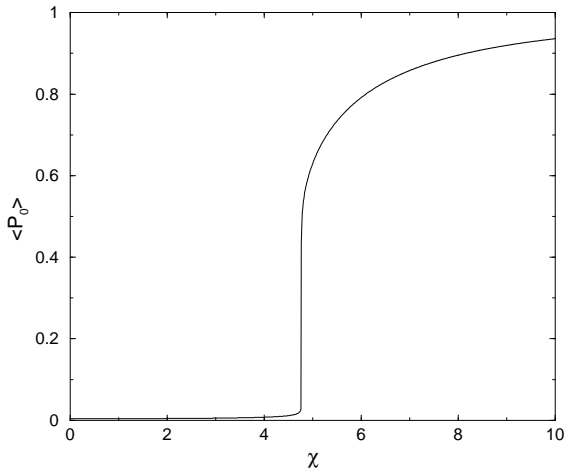}}
  \vspace{0.1cm}
  \caption{The time averaged probability of finding the particle at the
           impurity site 0 (i.e., $ < P_{0} >$) of the Cayley tree of $K=2$ 
           plotted against $\chi$. Sharp transition at $\chi=4.77$ is found.}
  \label{fig4}
  \vspace{0.2cm}
  \end{center}
\end{figure}

The time averaged probability is also plotted in Fig.~4 
as a function of the nonlinear parameter $\chi$. 
The sharp transition of the $< P_{0} >$ (the time averaged probability 
at the impurity site) at $\chi \simeq 4.77$ is also clear from the Fig.~4.

The self-trapping transition for the $K=1$ case, i.e., for a single 
nonlinear impurity embedded in a one-dimensional lattice has been 
studied in detail by Dunlap {\em et al.}\cite{21},
and the transition is found at $\chi=3.205$. 
However, we notice that the self-trapping transition for a sinlge
impurity embedded in the Cayley tree with $K=2$ is sharper and clearer
when compared with the case for the one-dimensional chain.

In order to observe the dependence of the critical value of $\chi$ as a 
function of the connectivity $K$ of the Cayley tree, we plotted in Fig.~5
$\chi_{cr}$ for various values of $K \ge 2$ on a double logarithmic scale.
The data points lie surprisingly well on the straight line,
implying that the critical value obeys the power-law
\begin{equation}
\chi_{cr} = \alpha ~ K^\beta ,
\end{equation}
with $\alpha \simeq 3.41$ and $\beta \simeq 0.484$.
It should be noted that the critical value $\chi_{cr} \simeq 3.41$ for $K=1$ 
is prominently different from the critical value for a one-dimensional system.
This implies that the power-law is valid only for $K \ge 2$. 
We therefore conclude that the geometry of the host lattice results in 
different critical values of $\chi$ for different values of $K \geq 2$,
obeying the power-law behavior in Eq.~(9).

\vspace{0.7cm}
\begin{figure}
  \begin{center}
  \leavevmode
  \epsfxsize=7.5cm
  \centerline{\epsfbox{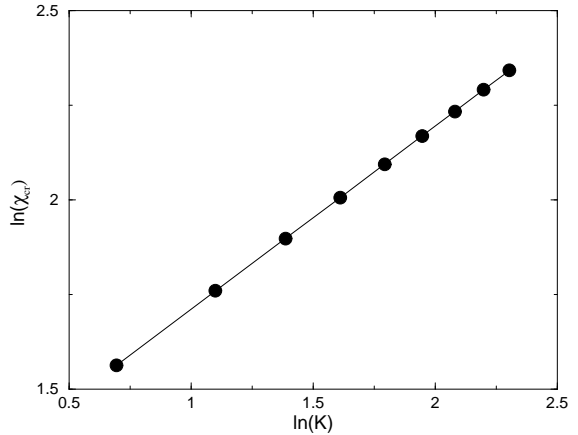}}
  \vspace{0.2cm}
  \caption{The critical values of $\chi$ for self-trapping transition 
           in a Cayley tree with a single nonlinear impurity is plotted 
	   as a function of the connectivity of the Cayley tree.}
  \label{fig5}
  \end{center}
\end{figure}

\begin{figure}
  \begin{center}
  \leavevmode
  \epsfxsize=7.5cm
  \centerline{\epsfbox{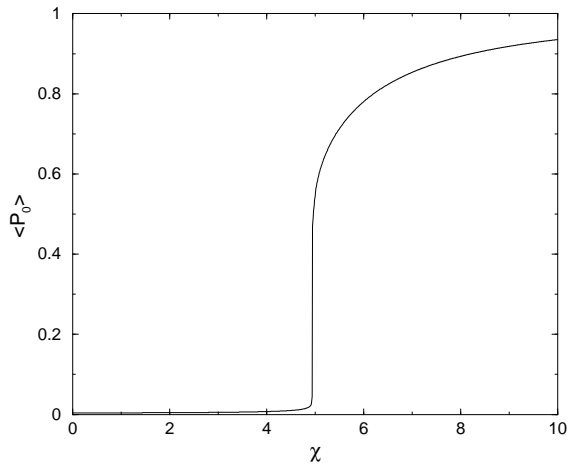}}
  \vspace{0.1cm}
  \caption{The time averaged probability of finding the paticle
           at the zeroth impurity site on a Cayley tree of $K=2$ 
	   with a dimeric nonlinear impurity plotted as a function of 
	   the nonlinearity parameter $\chi$.
	   Sharp transition is observed at $\chi = 4.95$}
  \label{fig6}
  \vspace{0.2cm}
  \end{center}
\end{figure}

We now consider the dimeric impurity embedded in the Cayley tree with
connectivity $K=2$. 
The particle is initially populated at one of the impurity sites. 
The probability of finding the particle at the initially populated site 
is calculated by solving the Eq.~(7) for various values of $\chi$,
and the time averaged result is plotted  in Fig.~6. 
The sharp transition is found at $\chi_{\rm cr} =4.95$. 
One interesting observation found from Fig.~6 is that there is no precursor 
(peak) in the $< P_{0} >$ before the permanent self-trapping transition occurs
at $\chi \simeq 4.95$, unlike the case in the one-dimensional system
with two impurities (Fig.~5 in Ref.\cite{22}) 
for which a peak is observed at $\chi \simeq 3.2$ just before the permanent
transition occurs at $\chi \simeq 4.22$.
Thus, the particle in the Cayley tree is always influenced by both the 
impurities present in the host whereas the particle does not feel 
the presence of the second impurity in the one-dimensional system 
below $\chi \simeq 3.2$. 
>From Eq.~(7) we note that the Cayley tree of connectivity $K$ 
with dimeric impurity reduces to a one-dimensional chain 
with the hopping element between the impurity sites
reduced from untiy to $V_{\rm eff} = \frac{1}{\sqrt{K}}$.
For the Cayley tree $K \geq 2$, the hopping element between the impurity
sites in the transformed one-dimensional system becomes less than
or equal to $\frac{1}{\sqrt{2}}$.
Since the peak before the permanent transition disappears in the
case of Cayley tree, we suspect that there must be a critical value
for the hopping element (say, $V_{\rm eff}^{\rm cr}$) between the
dimeric impurity sites of a one-dimensional chain while keeping the
other hopping elements unity such that the peak disappears for 
$V_{\rm eff} \leq V_{\rm eff}^{\rm cr}$.
In order to verify this, we perform the numerical calculation of the
time averaged probability for the particle to be at the initially 
populated impurity site of a one-dimensional chain with a dimeric 
nonlinear impurity for various values of $V_{\rm eff}$ 
(the hopping element between the impurity sites)
while keeping the other elements unity. 
We interestingly observed that the gap ($\Delta\chi$) between the peak 
and the permanent transition point reduces as the $V_{\rm eff}$ 
decreases and eventually vanishes at a critical value of 
$V_{\rm eff} = \frac{1}{\sqrt 2}$ (as shown in Fig. 7), i.e., 
the peak before the permanent transition dieappears 
when the system is equivalent to a Cayley tree with connectivity equal to 2. 
This implies that the structure of Cayley tree is responsible for the 
absence of the peak before the permanent transition.

\vspace{0.7cm}
\begin{figure}
  \begin{center}
  \leavevmode
  \epsfxsize=7.5cm
  \centerline{\epsfbox{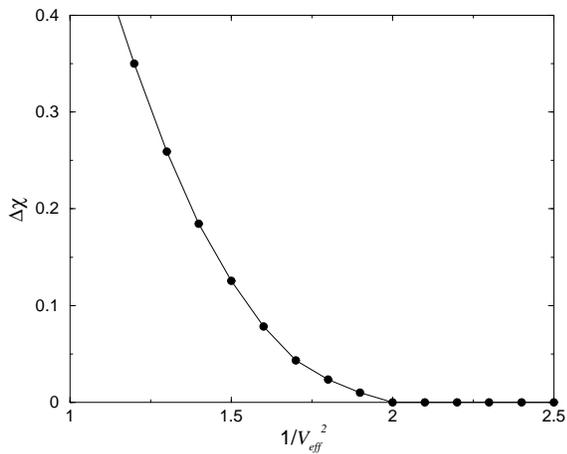}}
  \vspace{0.1cm}
  \caption{Plot of the gap $\Delta \chi$ between the peak and the permanent 
           transition point in the time averaged probability of the
	   particle at the initially populated impurity site in a
	   one-dimensional chain with a dimeric nonlinear impurity 
	   against the inverse of the square of the hopping element
	   between the impurity sites.
	   Other hopping elements are set to be unity.}
  \label{fig7}
  \vspace{0.2cm}
  \end{center}
\end{figure}

The critical value of $\chi$ for various values of the connectivity of 
the Cayley tree is calculated and found to obey the same power-law
behavior as in Eq.~(9), but with different values of $\alpha$ and $\beta$,
i.e., $\alpha \simeq 3.554$ and $\beta \simeq 0.465$.
We note that the critical value of $\chi$ for the self-trapping 
increases due to the presence of one more impurity. 
For a one-dimensional system the critical value is found to be 4.22 which
deviates prominently from the value estimated from Eq.~(9) 
with the values of $\alpha$ and $\beta$ mentioned above and $K=1$.
This is again due to the different geometry of the host lattice.

We now consider the perfectly nonlinear Cayley tree. 
The particle is initially placed at any arbitrary site. 
The sharp self-trapping transition is also observed in this case 
for various values of $K$. 
The critical value of the nonlinear parameter $\chi_{cr}$ 
for self-trapping transition again obeys the same power-law behavior as
in Eq.~(9) but with $\alpha \simeq 3.766$ and $\beta \simeq 0.445$. 
Here we find that the critical values are larger than those for the case of a 
single and dimeric impurity; however, the difference is not appreciably large.
Therefore, it appears that the self-trapping transition is influenced 
by only few nonlinear neighbors around the initially populated impurity site.

Finally it is worthy to mention that the formation of stationary localized
state in the Cayley tree due to a single nonlinear impurity, dimeric 
nonlinear impurity and also in a perfectly nonlinear Cayley tree
has been studied by Gupta and Kundu\cite{Gupta+Kundu:1997,Kundu+Gupta:1998}
earlier and in the above three cases the bifurcations in the 
stationary states were observed which in turn support the 
occurrence of the self-trapping transitions in the work.

\section{summary}
The self-trapping transition due to a single and a dimeric nonlinear
impurity embedded in the Cayley tree is studied. 
Furthermore, the self-trapping transition in a perfectly nonlinear 
Cayley tree is observed. 
Very sharp self-trapping transition is observed for all systems considered here.
The geometry of the host lattice is responsible for such a sharp transition.
The critical value of the self-trapping transition increases 
as the number of nonlinear impurities in the host lattice increases. 
The critical value of $\chi$ is found to obey a power law
against the connectivity of the Cayley tree for all cases. 
Results are compared with those for the one-dimensional system.

\acknowledgements
This work was supported by the Postdoc program at Kyungpook National
University in the year 2000. 
Work done by SBL is partially supported by the Korea Science and 
Engineering Foundation under Grant No.~KOSEF 981-0207-029-2.

\end{document}